# TOWARDS AMELIORATING CYBERCRIME AND CYBERSECURITY

**Azeez Nureni Ayofe**

Department of Computer Science,

College of Natural and Applied Sciences,

Fountain University, Osogbo,

Osun State, Nigeria.

E-mail address: nurayhn@yahoo.ca

**Osunade Oluwaseyifunmitan**

Department of Computer Science,

University of Ibadan, Nigeria.

seyitano@yahoo.com

o.osunade@mail.ui.edu.ng

**ABSTRACT--** Cybercrime is becoming ever more serious. Findings from 2002 Computer Crime and Security Survey show an upward trend that demonstrates a need for a timely review of existing approaches to fighting this new phenomenon in the information age. In this paper, we provide an overview of Cybercrime and present an international perspective on fighting Cybercrime.

This work seeks to define the concept of cyber-crime, identify reasons for cyber-crime, how it can be eradicated, look at those involved and the reasons for their involvement, we would look at how best to detect a criminal mail and in conclusion, proffer recommendations that would help in checking the increasing rate of cyber-crimes and criminals.

1.**INTRODUCTION**

Over the past twenty years, unscrupulous computer users have continued to use the computer to commit crimes; this has greatly fascinated people and evoked a mixed feeling of admiration and fear. This phenomenon has seen sophisticated and unprecedented increase recently and has called for quick response in providing laws that would protect the cyber space and its users. The level of sophistication has gone high to the point of using the system to commit murder and other havoc. The first recorded cyber murder committed in the United States seven years ago according to the Indian Express, January, 2002 "has to do with an underworld don in hospital to undergo a minor surgery. His rival went ahead to hire a computer expert who altered his prescriptions through hacking the hospital's computer system. He was administered the altered prescription by an innocent nurse, this resulted in the death of the patient"[i]

This work gives a brief overview of cyber-crime, explains why people are involved in cyber-crime, look at those involved and the reasons for their involvement, we would look at how best to detect a criminal mail and in conclusion, proffer recommendations that would help in checking the increasing rate of cyber-crimes and criminals. These guides provide general outlines as well as specific techniques for implementing cyber security.

2. **METHODOLOGY**

This study was carried out purposely to explain clearly the concept of Cybercrime and Cybersecurity and provide adequate and sufficient ways of getting out of these problems in the present days of internet usage and applications.

The instruments used were questionnaires, personal interviews, observation, and information on the internet as well as report from both radio and electronic media.

The authors conducted personal interviews with twenty two internet users to gather their views on the causes and their experiences with Cybercrime and Cybersecurity. In addition, fifty three questionnaires were distributed to the following categories of internet users: bankers, students, directors and university lecturers with aim of seeking their views and opinions on these issues.

Consequently, the information gathered through all the above instruments were analyzed and the approach towards ameliorating these phenomenon were proffered for both the government and corporate bodies for implementation.

3. **WHAT IS CYBER – CRIME?**

Cyber-crime by definition is any harmful act committed from or against a computer or network, it differs according to McConnell International, "from most terrestrial crimes in four ways: they are easy to learn how to commit, they require few resources relative to the potential damages caused, they can be committed in a jurisdiction without being physically present in it and fourthly, they are often not clearly illegal."[ii]

Another definition given by the Director of Computer Crime Research Centre (CCRC) during an interview on the 27th April, 2004, is that "cyber-crime ('computer crime') is any illegal behavior directed by means of electronic operations that targets the security of computer systems and the data processed by them."[iii] In essence, cyber-crime is crime committed in a virtual space and a virtual space is fashioned in a way that information about persons, objects, facts, events, phenomena or processes are represented in mathematical, symbol or any other way and transferred through local and global networks.

From the above, we can deduce that cyber crime has to do with wrecking of havoc on computer data or networks through interception, interference or destruction of such data or systems. It involves committing crime against computer systems or the use of the computer in committing crimes.

This is a broad term that describes everything from electronic cracking to denial of service attacks that cause electronic commerce sites to lose money. ". Mr. Pavan Duggal, who is the President of www.cyberlaws.net and consultant, in a report has clearly defined the various categories and types of cybercrimes. Cybercrimes can be basically divided into 3 major categories

1. Cybercrimes against persons.
2. Cybercrimes against property.
3. Cybercrimes against government.

   3.1  **Cybercrimes against persons**:
Cybercrimes committed against persons include various crimes like transmission of child-pornography, harassment of any one with the use of a computer such as e-mail. The trafficking, distribution, posting, and dissemination of obscene material including pornography and indecent exposure, constitutes one of the most important Cybercrimes known today. The potential harm of such a crime to humanity can hardly be amplified. This is one Cybercrime which threatens to undermine the growth of the younger generation as also leave irreparable scars and injury on the younger generation, if not controlled.

A minor girl in Ahmedabad was lured to a private place through cyberchat by a man, who, along with his friends, attempted to gangrape her. As some passersby heard her cry, she was rescued.

Another example wherein the damage was not done to a person but to the masses is the case of the Melissa virus. The Melissa virus first appeared on the internet in March of 1999. It spread rapidly throughout computer systems in the United States and Europe. It is estimated that the virus caused 80 million dollars in damages to computers worldwide.

In the United States alone, the virus made its way through 1.2 million computers in one-fifth of the country's largest businesses. David Smith pleaded guilty on Dec. 9, 1999 to state and federal charges associated with his creation of the Melissa virus. There are numerous examples of such computer viruses few of them being "Melissa" and "love bug".

Cyber harassment is a distinct Cybercrime. Various kinds of harassment can and do occur in cyberspace, or through the use of cyberspace. Harassment can be sexual, racial, religious, or other. Persons perpetuating such harassment are also guilty of cybercrimes. Cyberharassment as a crime also brings us to another related area of violation of privacy of citizens. Violation of privacy of online citizens is a Cybercrime of a grave nature. No one likes any other person invading the invaluable and extremely touchy area of his or her own privacy which the medium of internet grants to the citizen.

 3.2  **Cybercrimes against property**: The second category of Cyber-crimes is that of Cybercrimes against all forms of property. These crimes include computer vandalism (destruction of others' property), transmission of harmful programmes.

A Mumbai-based upstart engineering company lost a say and much money in the business when the rival company, an industry major, stole the technical database from their computers with the help of a corporate cyberspy.

3.3 **Cybercrimes against government:** The third category of Cyber-crimes relate to Cybercrimes against Government. Cyberterrorism is one distinct kind of crime in this category. The growth of internet has shown that the medium of Cyberspace is being used by individuals and groups to threaten the international governments as also to terrorize the citizens of a country. This crime manifests itself into terrorism when an individual "cracks" into a government or military maintained website.

In a report of expressindia.com, it was said that internet was becoming a boon for the terrorist organizations. According to Mr. A.K. Gupta, Deputy Director (Co-ordination), CBI, terrorist outfits are increasingly using internet to communicate and move funds. "Lashker-e-Toiba is collecting contributions online from its sympathizers all over the world. During the investigation of the Red Fort shootout in Dec. 2000, the accused Ashfaq Ahmed of this terrorist group revealed that the militants are making extensive use of the internet to communicate with the operatives and the sympathizers and also using the medium for intra-bank transfer of funds".

Cracking is amongst the gravest Cyber-crimes known till date. It is a dreadful feeling to know that a stranger has broken into your computer systems without your knowledge and consent and has tampered with precious confidential data and information.

Coupled with this the actuality is that no computer system in the world is cracking proof. It is unanimously agreed that any and every system in the world can be cracked. The recent denial of service attacks seen over the popular commercial sites like E-bay, Yahoo, Amazon and others are a new category of Cyber-crimes which are slowly emerging as being extremely dangerous.

Cyber crime can be broadly defined as criminal activity in which computer or computer networks are a tool, a target or a medium for the crime.

4. **Various types of cyber crimes include:**
   4.1 Unauthorized access of hosts- more commonly known as hacking. Hacking can take various forms, some of which might not always involve deep technical knowledge.
   - Social engineering involves "talking" your way into being given access to a computer by an authorized user.
   - A divide exists between individuals who illegally break into computers with malicious intent, or to sell information garnered from the compromised computer, known as "crackers" or black hats", and those who do it out of curiosity or to enhance their technical prowess- known as "hackers" or "white hats".

   4.2 Spamming – involves mass amounts of email being sent in order to promote and advertise products and websites.
   - Email spam is becoming a serious issue amongst businesses, due to the cost overhead it causes not only in regards to bandwidth consumption but also to the amount of time spent downloading/ eliminating spam mail.
   - Spammers are also devising increasingly advanced techniques to avoid spam filters, such as permutation of the emails contents and use of imagery that cannot be detected by spam filters.

   4.3 Computer Fraud/ "Phishing" scams- South Africa has recently been afflicted by an onset of intricate scams that attempt to divulge credit and banking information from online banking subscribers.
   - These are commonly called "Phishing' scams, and involve a level of social engineering as they require the perpetrators to pose as a trustworthy representative of an organization, commonly the victims bank.

   4.4 Denial of Service Attacks- Not to be confused with unauthorized computer access and hacking.
   - Denial of Service or DoS attacks involve large volumes of traffic being sent to a host or network, rendering it inaccessible to normal users due to sheer consumption of resources.
   - Distributed Denial of Service attacks involve multiple computers being used in an attack, in many

cases through the use of "zombie" servers, which are trojanised programs that attackers install on various computers.
- Often legitimate computer users have no idea they are involved in denial of service attacks due to the stealthy nature of the zombie software.

4.5 Viruses, Trojans and Worms- These three all fall into a similar category as they are software designed to "infect" computers- or install themselves onto a computer without the users permission, however they each operate very differently.
- Many computer users have experienced the frustration of having a malicious virus wreck havoc upon their computers and data, but not all viruses have a malicious payload.
- Trojan is a program that allows for the remote access of the computer it's installed on. Trojans exist for multiple performs and have varying degrees in complexity.
- Worms make use of known vulnerabilities in commonly used software, and are designed to traverse through networks- not always with destructive ends, historically however worms have had devastating effects such as the infamous Code Red and Melissa worms.
- Intellectual Property Theft- Intellectual property theft in relation to cyber crime deals mainly with the bypassing of measures taken to ensure copyright- usually but not restricted to software.

4.6 Other types of cyber crime could be categorized under the following:
1. Unlawful access to computer information - 8002 crimes.
2. Creation, use and distribution of malware or machine carriers with such programs- 1079.
3. Violation of operation rules of computers, computer system or networks- 11.
4. Copyright and adjacent rights violation- 528.

## 5. **CAUSES OF CYBER – CRIME**

There are many reasons why cyber-criminals commit cyber-crime, chief among them are these three listed below:
- Cyber crimes can be committed for the sake of recognition. This is basically committed by youngsters who want to be noticed and feel among the group of the big and tough guys in the society. They do not mean to hurt anyone in particular; they fall into the category of the Idealists; who just want to be in spotlight.
- Another cause of cyber-crime is to make quick money. This group is greed motivated and is career criminals, who tamper with data on the net or system especially, e-commerce, e-banking data information with the sole aim of committing fraud and swindling money off unsuspecting customers.
- Thirdly, cyber-crime can be committed to fight a cause one thinks he believes in; to cause threat and most often damages that affect the recipients adversely. This is the most dangerous of all the causes of cyber-crime. Those involve believe that they are fighting a just cause and so do not mind who or what they destroy in their quest to get their goals achieved. These are the cyber-terrorists.

## 6. **HOW TO ERADICATE CYBER – CRIME**

Research has shown that no law can be put in place to effectively eradicate the scourge of cyber-crime. Attempts have been made locally and internationally, but these laws still have shot-comings. What constitutes a crime in a country may not in another, so this has always made it easy for cyber criminals to go free after being caught.

These challenges notwithstanding, governments should in the case of the idealists, fight them through education not law. It has been proven that they help big companies and government see security holes which career criminals or even cyber-terrorist could use to attack them in future. Most often, companies engage them as consultants to help them build solid security for their systems and data. "The Idealists often help the society: through their highly mediatised and individually harmless actions, they help important organizations to discover their high-tech security holes…."[iv] The enforcement of law on them can only trigger trouble, because they would not stop but would want to defy the law. " Moreover, if the goal of the cyber-crime legislation is to eradicate cyber-crime, it mint well eradicate instead a whole new culture...."[v] Investments in education is a much better way to prevent their actions.

Another means of eradicating cyber-crime is to harmonize international cooperation and law, this goes for the greed motivated and cyber-terrorists. They can not be fought by education, because they are already established criminals, so they can not behave. The only appropriate way to fight them is by enacting new laws, harmonize international legislations and encourage coordination and cooperation between national law enforcement agencies.

### 7. **WHO ARE INVOLVED**

Those involved in committing cyber-crimes are in three categories and they are:

- 7.1 **THE IDEALISTS (Teenager).** They are usually not highly trained or skilful, but youngsters between the ages of 13 – 26 who seek social recognition. They want to be in the spotlight of the media. Their actions are globally damageable but individually negligible. "Like denying a lot of important e-commerce servers in February, 2000 is said to have caused high damages to these companies."[vi] Most often they attack systems with viruses they created; their actual harm to each individual is relatively negligible. By the age of 26 to 26 when they have matured and understood the weight of their actions, they lose interest and stop.

- 7.2 **THE GREED – MOTIVATED (Career Criminals).** This type of cyber-criminals is dangerous because they are usually unscrupulous and are ready to commit any type of crime, as long as it brings money to them. "They started the child pornography often called cyber-pornography which englobes legal and illegal pornography on the internet." [vii] They are usually very smart and organized and they know how to escape the law enforcement agencies. These cyber-criminals are committing grievous crimes and damages and their unscrupulousness, particularly in child-pornography and cyber-gambling is a serious threat to the society. Example to show how serious a threat they pose to the society is "the victim of the European bank of Antigua are said to have lost more than $10million"[viii] "...theft of valuable trade secrets: the source code of the popular micro-soft windows exploration system by a Russian based hacker could be extremely dangerous... the hackers could use the code to break all firewalls and penetrated remotely every computer equipped with windows were confirmed. Another usage could be the selling of the code to competitors."[ix]

- 7.3 **THE CYBER – TERRORISTS.** They are the newest and most dangerous group. Their primary motive is not just money but also a specific cause they defend. They usually engage in sending threat mails, destroying data stored in mainly government information systems just to score their point. The threat of cyber-terrorism can be compared to those of nuclear, bacteriological or chemical weapon threats. This disheartening issue is that they have no state frontiers; can operate from any where in the world, and this makes it difficult for them to get caught. The most wanted cyber-terrorist is Osama Bin Laden who is said to "use stegranography to hide secret messages within pictures, example, a picture of Aishwarya Rai hosted on the website could contain a hidden message to blow up a building."[x] A surprising fact is that these hidden messages do not alter the shape, size or look of the original pictures in any way.

### 8. **A CRIMINAL MAIL**

Another type of Cybercrime which is being currently researched on but not as popular as those stated above is a criminal mail.

A criminal mail is usually sent to networks with the aim of either corrupting the system or committing fraud. The way to detect such mails is by putting security measures in place which would detect criminal patterns in the network. News Story by Paul Roberts, of IDG News Service says that Unisys Suite has a system called the "Unisys Active Risk Monitoring System (ARMS) which helps banks and other organizations spot patterns of seemingly unrelated events that add up to criminal activity."[xi] Actimize Technology Ltd based in New York has developed technology that enables organizations to do complex data mining and analysis on stored information and transaction data without needing to copy it to a separate data warehouse. "The actimize software runs on the Microsoft Corp. Windows NT or Windows 2002 platform and can be developed on standard server hardware with either four to eight processors, Katz said."[xii]

Eric J. Sinrod in his article 'What's Up With Government Data Mining' states that the United States "Federal Government has been using data mining techniques for various purposes, from attempting to improve service to trying to detect terrorists patterns and activities."[xiii] The most effective way to detect criminal mails is to provide security gadgets, educate employees on how to use them, and to be at alert for such mails, above all, making sure no security holes is left unattended to.

The world over Cybercrime has taken deep root and the use of cyberspace by sophisticated cyber criminals has assumed serious portion today. Criminals and terrorists associated with drug trafficking, terrorist outfits are employing internet for anti social, anti national and criminal activities with impunity.

Terrorist groups are deftly using internet for passing on information with regard to executing various terrorist acts having serious negative impact on human life. The cyber-terrorists have even acquired the capability to penetrate computer systems using "logic bombs" (coded devices that can be remotely detonated), electro magnetic pulses and high-emission radio frequency guns, which blow devastating electronic wind through a computer system. The hackers have gone to the extent of distributing free hacking software—Rootkit, for instance—to enable an intruder to get root access to a network and then control as though they were the system's administrators.

Cyber crime levels are on the rise in Nigeria, examples of large scale cyber crimes over the past few years include:
- The phishing scams that have recently afflicted many of Nigeria's larger banks and their clients.
- Key logging software that was able to capture banking and other details of online bankers.

Statistical research performed in the UK revealed that cyber crime and software flaws were costing Britain up to £10 billion in losses annually.
According to the survey 50% of all businesses (**source:** The Law and Technology of Software Security: Nigerian Cybercrime Working Group (NCWG) 8th Nigerian Software Exhibition - NISE 2004) were affected by cyber crime, showing a giant increase in cyber crime occurrences in the UK when compared to a 2000 survey which revealed that only 25% of respondents had been cyber crime victims.

**Table1** showing Internet Usage and World internet usage and population statistics as at March 30, 2009.

| World Region | Population (2008 Est.) | Population % of world | Internet usage, Latest Data | % population (penetration) | Usage % of world | Usage Growth 2008 |
|---|---|---|---|---|---|---|
| Africa | 941,249,130 | 14.2% | 44,361,940 | 4.7% | 3.4% | 882.7% |
| Asia | 3,733,783,474 | 56.6% | 510,478,743 | 13.7% | 38.7% | 346.6% |
| Europe | 801,821,187 | 12.1% | 348,125,847 | 43.4% | 26.4% | 231.2% |
| Middle East | 192,755,045 | 2.9% | 33,510,500 | 17.4% | 2.5% | 920.2% |
| North America | 334,659,631 | 5.1% | 238,015,529 | 71.1% | 18.0% | 120.2% |
| Latin America/Caribbean | 569,133,474 | 8.6% | 126,203,714 | 22.2% | 9.6% | 598.5% |
| Australia | 33,569,718 | 0.5% | 19,175,836 | 57.1% | 1.5% | 151.6% |
| World Total | 6,606,971,659 | 100.0% | 1,319,872,109 | 20.0% | 100.0% | 265.6% |

8.1 **Technology Viewpoint**
- Advances in high-speed telecommunications, computers and other technologies are creating new opportunities for criminals, new classes of crimes, and new challenges for law enforcement.

8.2 **Economy Viewpoint**
- Possible increases in consumer debt may affect bankruptcy filings.

- Deregulation, economic growth, and globalization are changing the volume and nature of anticompetitive behaviour.
- The interconnected nature of the world's economy is increasing opportunities for criminal activity.

### 8.2 **Government Viewpoint**
- Issues of criminal and civil justice increasingly transcend national boundaries, require the cooperation of foreign governments, and involve treaty obligations, multinational environment and trade agreements and other foreign policy concerns.

### 8.3 **Social-Demographic Viewpoint**
- The numbers of adolescents and young adults, now the most crime-prone segment of the population are expected to grow rapidly over the next several years.

### 8.4 **Computer as an instrument facilitating crime**

Computer is used as an instrument facilitating crime. Most vivid example of computers being used as an instrument of Cybercrime has been the recent attack on parliament where computer and internet was used in a variety of ways to perpetrate the crime. The terrorist and criminals are using internet methods such as e-mail, flash encrypted messages around the globe. Frauds related to electronic banking or electronic commerce are other typical examples. In these crimes, computer programmes are manipulated to facilitate the crimes namely,

a. Fraudulent use of Automated Teller Machine (ATM) cards and accounts;
b. Credit card frauds;
c. Frauds involving electronic finds transfers;
d. Telecommunication Frauds; and
e. Frauds relating to Electronic Commerce and Electronic Data Interchange.

The information technology (IT) infrastructure which is now vital for communication, commerce, and control of our physical infrastructure, is highly vulnerable to terrorist and criminal attacks. The private sector has an important role in securing the Nation's IT infrastructure by deploying sound security products and adopting good security practices.

But the Federal government also has a key role to play by supporting the discovery and development of cyber security technologies that underpin these products and practices.

Improving the Nation's cyber security posture requires highly trained people to develop, deploy, and incorporate new cyber security products and practices. The number of such highly trained people is too small given the magnitude of the challenge. The situation has been exacerbated by the insufficient and unstable funding levels for long-term, civilian cyber security research, which universities depend upon to attract and retain faculty.

### 8.5 **Software vulnerability**
Network connectivity provides "door-to_door" transportation for attackers, but vulnerabilities in the software residing in computers substantially compound the cyber security problem. The software development methods that have been the norm fail to provide the high-quality, reliable, and secure software that the IT infrastructure requires. Software development is not yet a science or a rigorous discipline, and the development process by and large is not controlled to minimize the vulnerabilities that attackers exploit. Today, as with cancer, vulnerable software can be invaded and modified to cause damage to previously healthy software, and infected software can replicate itself and be carried across networks to cause damage in other systems.

Like cancer, these damaging processes may be invisible to lay person even though experts recognize that their threat is growing. And as in cancer, both preventive actions and research are critical, the former to minimize damage today and the latter to establish a foundation of knowledge and capabilities that will assist the cyber security professionals of tomorrow reduce the risk and minimize damage for the long term.

*8.6* ***Domestic and international law enforcement***. A hostile party using an Internet-connected computers thousands of miles away can attack an Internet- connected computers in the United States as easily as if he or she were next door. It is often difficult to identify the perpetrator of such an attack, and even when a perpetrator is identified, criminal prosecution across national boundaries is problematic.

*8.7* ***Education.*** We need to educate citizens that if they are going to use the internet, they need to continually maintain and update the security on their system so that they cannot be compromised, for example, to become agents in a DDoS attack or for "spam" distribution. We also need to educate corporations and organizations in the best practice for effective security management. For example, some large organizations now have a policy that all systems in their purview must meet strict security guidelines. Automated updates are sent to all computers and servers on the internal network, and no new system is allowed online until it conforms to the security policy.

*8.8 **Information security.*** Information security refers to measures taken to protect or preserve information on a network as well as the network itself.
The alarming rise of premeditated attacks with potentially catastrophic effects to interdependent networks and information systems across the globe has demanded that significant attention is paid to critical information infrastructure protection initiatives.

For many years governments have been protecting strategically critical infrastructures, however in recent times the information revolution has transformed all areas of life. The way business is transacted, government operates, and national defence is conducted has changed. These activities now rely on an interdependent network of information technology infrastructures and this increases our risk to a wide range of new vulnerabilities and threats to the nation's critical infrastructures. These new cyber threats are in many ways significantly different from the more traditional risks that governments have been used to addressing. Exploiting security flaws appears now to be far easier, less expensive and more anonymous than ever before.

The increasing pervasiveness, connectivity and globalization of information technology coupled with the rapidly changing, dynamic nature of cyber threats and our commitment to the use of ICT for socio-economic development brings about the critical need to protect the critical information infrastructures to provide
greater control. This means that governments must adopt an integrated approach to protect these infrastructures from cyber threats.

9. **RESULTS AND DISCUSSION**

The study revealed that three categories of people are involved in committing Cybercrime (The idealists, the Greed-motivated and the cyber-terrorist). It was equally gathered that these categories of people have contributed in no small measure to cyber terrorism.

During the course of interview, it was learnt that four out of twenty two people interviewed were victims of Cybercrime and seven others have their relatives affected in one way or the other.

It is equally obvious that that Cybercrime committed against person, property and government have claimed millions of US Dollar[v] and has affected up to 56% of e-commerce globally[iv].

Against this backdrop, the authors offered the recommendations in this paper as panacea for Cybercrime and Cybersecurity with a view of having a reliable and consistent internet usage in the world.

10. **CONCLUSION**
Cybercrime and Cyber security has become a subject of great concern to all governments of the world. Nigeria, representing the single largest concentration of people of Africa decent has an important role to play. This situation has almost reached an alarming point, according to various studies and countries which neglects and /fail to respond timely and wisely, will pay very dearly for it.

It has been deduced from this study that reliance on terrestrial laws is still an untested approach despite progress being made in many countries, they still rely on standard terrestrial laws to prosecute cyber crimes and these laws are archaic statutes that have been in existence before the coming of the cyberspace.  Also weak penalties limit deterrence: countries with updated criminal statutes still have weak penalties on the criminal statutes; this can not deter criminals from committing crimes that have large-scale economic and social effect on the society.  Also a global patchwork of laws creates little certainty; little consensus exist among countries regarding which crimes need to be legislated against.  Self protection remains the first line of defense and a model approach is needed by most countries; especially those in the developing world looking for a model to follow.  They recognize the importance of outlawing malicious computer-related acts in a timely manner or in order to promote a secure environment for e-commerce.

Cyber-crime with its complexities has proven difficult to combat due to its nature. Extending the rule of law into the cyberspace is a critical

*(IJCSIS) International Journal of Computer Science and Information Security,*
*Vol. 3, No. 1, 2009*step towards creating a trustworthy environment for people and businesses. Since the provision of such laws to effectively deter cyber-crime is still a work in progress, it becomes necessary for individuals and corporate bodies to fashion out ways of providing security for their systems and data. To provide this self-protection, organizations should focus on implementing cyber-security plans addressing people, process and technology issues, more resources should be put in to educate employees of organizations on security practices, "develop thorough plans for handling sensitive data, records and transactions and incorporate robust security technology- -such as firewalls, anti-virus software, intrusion detection tools and authentication services--."[xiv]

11. **RECOMMENDATION**

By way of recommendations, these kinds of actions (both in form of security, education and legislation) are suggested following the weak nature of global legal protection against cyber crime:

**A. Legislation approach:**

- Laws should apply to cyber-crime— National governments still are the major authority who can regulate criminal behavior in most places in the world. So a conscious effort by government to put laws in place to tackle cyber-crimes would be quite necessary.
- Review and enhance Nigeria cyber law to address the dynamic nature of cyber security threats;
- Ensure that all applicable local legislation is complementary to and in harmony with international laws, treaties and conventions;
- Establish progressive capacity building programmes for national law enforcement agencies;
- There should be a symbiotic relationship between the firms, government and civil society to strengthen legal frameworks for cyber-security. An act has to be crime in each jurisdiction before it can be prosecuted across a border. Nation must define cyber-crimes in similar manner, to enable them pass legislation that would fight cyber-crimes locally and internationally.

**B. Security approach**

- Strengthening the trust framework, including information security and network security, authentication, privacy and consumer protection, is a prerequisite for the development of the information society and for building confidence among users of ICTs;
- A global culture of Cyber security needs to be actively promoted, developed and implemented in cooperation with all stakeholders and international expert bodies;
- Streamlining and improving the co-ordination on the implementation of information security measures at the national and international level;
- Establishment of a framework for implementation of information assurance in critical sectors of the economy such as public utilities, telecommunications, transport, tourism, financial services, public sector, manufacturing and agriculture and developing a framework for managing information security risks at the national level;
- Establishment of an institutional framework that will be responsible for the monitoring of the information security situation at the national level, dissemination of advisories on latest information security alerts and management of information security risks at the national level including the reporting of information security breaches and incidents;
- Promote secure e-commerce and e-government services;
- Safeguarding the privacy rights of individuals when using electronic communications and
- Develop a national cyber security technology framework that specifies cyber security requirement controls and baseline for individual network user;
- Firms should secure their network information. When organization provides security for their networks, it becomes possible to enforce property rights laws and punishment for whoever interferes with their property.

**C. Education/Research**

- Improving awareness and competence in information security and sharing of best practices at the national level through the development of a culture of Cybersecurity at national level.
- Formalize the coordination and prioritization of cyber security research and development activities; disseminate vulnerability advisories and threat warnings in a timely manner.
- Implement an evaluation/certification programme for cyber security product and systems;
- Develop, foster and maintain a national culture of security standardize and coordinate Cybersecurity awareness and education programmes;